\documentclass[authorversion,sigconf,nonacm]{acmart}

\AtBeginDocument{%
  \providecommand\BibTeX{{%
    \normalfont B\kern-0.5em{\scshape i\kern-0.25em b}\kern-0.8em\TeX}}}

\copyrightyear{2024} 
\acmYear{2024} 
\setcopyright{acmlicensed}\acmConference[CHI '24]{Proceedings of the CHI
Conference on Human Factors in Computing Systems}{May 11--16,
2024}{Honolulu, HI, USA}
\acmBooktitle{Proceedings of the CHI Conference on Human Factors in
Computing Systems (CHI '24), May 11--16, 2024, Honolulu, HI, USA}
\acmDOI{10.1145/3613904.3642101}
\acmISBN{979-8-4007-0330-0/24/05}

\usepackage{_macros}

\newcommand{\name}{EmoWear}
\newcommand{\baseline}{Baseline}

\newcommand{\re}[1]{\textcolor{black}{#1}}

\begin{document}

\title{EmoWear: Exploring Emotional Teasers for Voice Message Interaction on Smartwatches}
\titlenote{To appear at ACM CHI '24.}


\author{Pengcheng An}
\affiliation{%
  \institution{School of Design, SUSTech}
  \streetaddress{}
  \city{Shenzhen}
  \country{China}}
\email{anpc@sustech.edu.cn}

\author{Jiawen Zhu}
\author{Zibo Zhang}
\affiliation{%
  \institution{University of Waterloo}
  \streetaddress{}
  \city{Waterloo}
  \state{Ontario}
  \country{Canada}}
\email{jiawen.zhu@uwaterloo.ca}
\email{selenazhang131@gmail.com}

\author{Yifei Yin}
\affiliation{%
  \institution{University of Toronto Scarborough}
  \streetaddress{}
  \city{Scarborough}
  \state{Ontario}
  \country{Canada}}
\email{yifei.yin@mail.utoronto.ca}

\author{Qingyuan Ma}
\affiliation{%
  \institution{Chalmers University of Technology}
  \streetaddress{}
  \city{Gothenburg}
  \country{Sweden}}
\email{qingyuan0102@gmail.com}

\author{Che Yan}
\author{Linghao Du}
\affiliation{%
  \institution{Human-Machine Interaction Lab, Huawei Canada}
  \streetaddress{}
  \city{Markham}
  \state{Ontario}
  \country{Canada}}
\email{shino.yan@huawei.com}
\email{linghao.du@huawei.com}

\author{Jian Zhao}
\authornote{Corresponding Author.}
\affiliation{%
  \institution{University of Waterloo}
  \streetaddress{}
  \city{Waterloo}
  \state{Ontario}
  \country{Canada}}
\email{jianzhao@uwaterloo.ca}



\renewcommand{\shortauthors}{P. An, J. Zhu, Z. Zhang, Y. Yin, Q. Ma, C. Yan, L. DU \& J. Zhao}

\begin{abstract}
Voice messages, by nature, prevent users from gauging the emotional tone without fully diving into the audio content. This hinders the shared emotional experience at the pre-retrieval stage. Research scarcely explored “Emotional Teasers”—pre-retrieval cues offering a glimpse into an awaiting message’s emotional tone without disclosing its content. We introduce EmoWear, a smartwatch voice messaging system enabling users to apply 30 animation teasers on message bubbles to reflect emotions. EmoWear eases senders’ choice by prioritizing emotions based on semantic and acoustic processing. EmoWear was evaluated in comparison with a mirroring system using color-coded message bubbles as emotional cues (N=24). Results showed EmoWear significantly enhanced emotional communication experience in both receiving and sending messages. The animated teasers were considered intuitive and valued for diverse expressions. Desirable interaction qualities and practical implications are distilled for future design. We thereby contribute both a novel system and empirical knowledge concerning emotional teasers for voice messaging.
\end{abstract}



\begin{teaserfigure}
  \includegraphics[width=\textwidth]{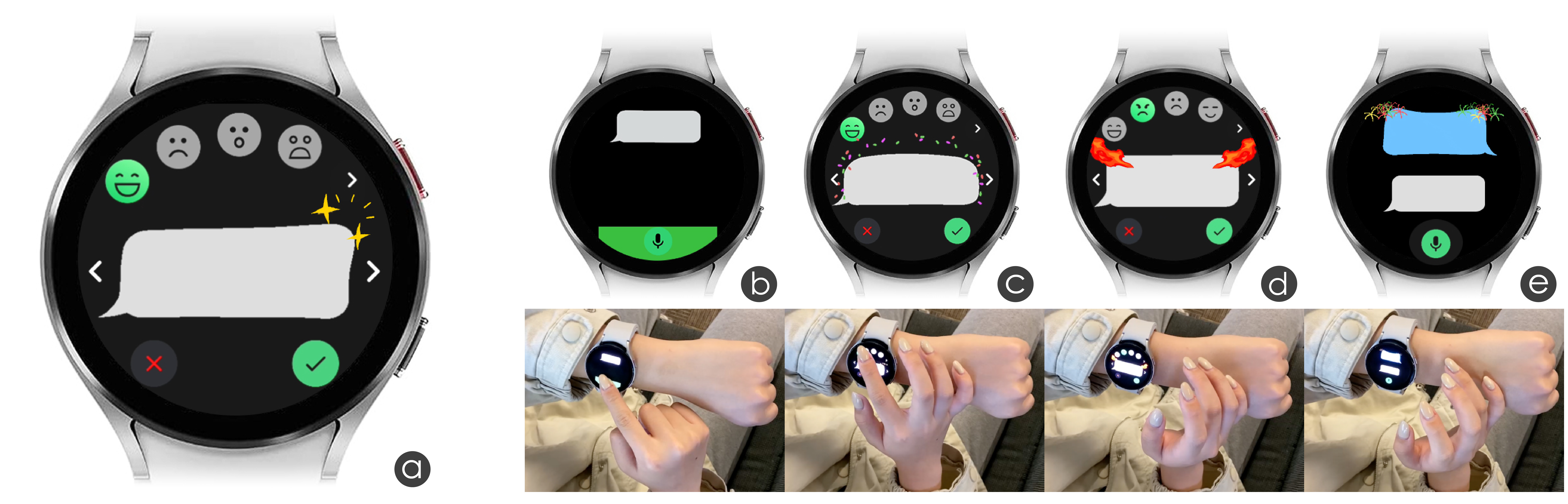}
  \vspace{-7mm}
  \caption{(a) \name{} is a smartwatch voice messaging system enabling users to apply 30 animated emotional teasers: pre-retrieval cues offering a glimpse into an awaiting message’s emotional tone before diving into the audio content; \name{} interaction flow typically includes: (b) recording the message, (c,d) browsing and selecting emotional teasers, (e) receiving the message.}
  \Description{}
  \label{fig:teaser}
\end{teaserfigure}

\maketitle

\section{Introduction}

Voice messages have become an integral component of our everyday communication, offering a personal touch through the asynchronous exchange of audio snippets.
Recent data suggest a surge in their usage. According to a YouGov poll conducted for Vox\footnote{https://www.vox.com/technology/23665101/voice-message-whatsapp-apple-text}, approximately 62\% of Americans have used voice messaging, with around 30\% using it weekly or more frequently. The trend is even more salient among younger individuals: 43\% of respondents aged 18 to 29 leverage voice messaging at least weekly. As reported by WhatsApp alone, over seven billion voice messages are sent by its users each day \footnote{https://blog.whatsapp.com/making-voice-messages-better}. 

The ubiquity of voice messaging is underscored by its support across various devices, notably including smartwatches. 
Given the limitation of reading or typing on the small watch screens, voice messaging stands out as a meaningful alternative to text-based messaging, especially when users are in nomadic scenarios or in the midst of other activities (where almost half of the voice message interactions took place \cite{UsageVoiceMessaging}). While voice messages ease communication on smartwatches, they also shift some effort from senders to receivers: their retrieval demands more effort and can be restricted in noisy surroundings or specific social occasions \cite{UsageVoiceMessaging}. When the retrieval has to be deferred, awaiting voice messages do not afford receivers a sneak peek at their emotional undertones.

In face-to-face interactions, we naturally rely on nonverbal social-emotional cues, such as facial expressions or body language \cite{vinciarelli2009socialSignal}, which remain visible throughout our conversation. In contrast, with pre-retrieval voice messages, the emotional information is not discernible until they are listened to.
Much like a black box, the emotional undertones a voice message remain hidden until ``opened''---in this case, with the voice recording played or transcribed into texts.

Visualizing emotional cues for pre-retrieval voice messages, therefore, could grant users an ``\textit{Emotional Teaser}'' before fully diving into the audio content. This could facilitate the experience of shared emotional understanding at the pre-retrieval stage, as well as setting up affective anticipation for the communication.
Recent research in Human-Computer Interaction (HCI) has just started to explore such emotional teasers for voice messaging on mobile phones. Chen et al.'s work \cite{bubbleColoring} pioneered this space by using colored voice message bubbles to indicate excitement, anger, sadness, and serenity. 
As related, the EmoBalloon system \cite{emoBalloon} depicts the arousal level detected in a text message through a generated explosion-shaped text bubble. 
Despite the original usage for text messaging, the above work shows the promise of utilizing message bubbles for emotional teasers of voice messages.

In smartwatch-based communication, 
little empirical knowledge has been accumulated regarding how senders and receivers of voice messages would experience such emotional teasers.
Nonetheless, in other forms of smartwatch-based social interaction, animations have been found as a preferred affective medium. Namely, in the works of Animo \cite{AnimoLiu} and Significant Otter \cite{Otter}, Liu et al. explored using abstract or elaborate animations for wearers to share their affective states based on bio-data. Their studies showed the benefits of animations in communicating nuanced affects between wearers, implying the prospect of using animations as emotional teasers for voice messages.

Motivated by these studies, the present work tackles the unaddressed opportunity of emotional teasers for voice messages on smartwatches. As prior cases separately surfaced the affordance of message bubbles \cite{emoBalloon,bubbleColoring} and the value of animations as affective cues \cite{AnimoLiu,Otter}, 
combining the two, we set out to explore affective animations of message bubbles as emotional teasers for pre-retrieval voice messages.

In particular, this paper presents the design and evaluation of \name{}, a smartwatch-based system that enables users to send and receive voice messages with animated emotional teasers (\autoref{fig:teaser}).
The \name{} system includes an intuitive front-end user interface to send or receive voice messages, which enables senders to apply 30 teaser animations on message bubbles to reflect emotions from six categories: happiness, sadness, calmness, fear, surprise, and anger. 
Moreover, \name{} is equipped with a back-end algorithm that incorporates a fusion model to process both the semantic and acoustic features of the input audio and outputs an emotion classification result.
Accordingly, the \name{} interface re-arranges the display order of the emotion categories to prioritize two probable ones to ease senders' selection while offering freedom for them to browse other options. 

Utilizing \name{} as a novel system to study about, as well as an inquiry tool to study with, we aim to answer the twofold research questions: 
\textbf{RQ1:} How users would experience message bubble animations as emotional teasers on smartwatches?
\textbf{RQ2:} What would be the relevant design opportunities and implications for emotional teasers of voice messages?
To concretely understand its potential, in our evaluation with 12 pairs of participants, we contrasted \name{} with the previously studied approach---using message bubble colors as emotion indicators---by developing a counterpart version of the \name{} interface (see \autoref{fig:baseline}). 
Our findings suggest that the animated message bubbles as emotional teasers, showed advantages in enhancing users' communication experience, helping senders express their emotions, and facilitating receivers to interpret the emotions in awaiting voice messages; and the pre-retrieval interpretation was perceived as aligned with the message content after accessing the audio.
Moreover, the qualitative findings contextualize how the \name{} emotional teasers enhanced the users' communication experience, offering valuable insights into their potential roles and benefits in daily communication contexts. 
Based on the findings, we generalize a list of preferred interaction qualities and contextual opportunities for future HCI design. We also discuss relevant implications for future HCI research to expand the knowledge and impacts of emotional teasers.

Our contribution is thereby twofold: (1) a smartwatch-based system that enables users to send and receive voice messages with emotional teasers and (2) an empirical understanding of how users would experience such emotional teasers, and relevant opportunities and implications for future HCI practice and research. 
\section{Related Work}
\subsection{Affective Enhancement in Text Messaging}
Although HCI research accumulated little knowledge about the design of emotional teaser features for voice messages, prior studies have extensively explored how various paralinguistic components could be employed as affective enhancement to accompany text messages and real-time audio transcripts. 

Emoticons, or emojis, is one of the most studied methods in the domain of text messaging. These symbols have a rich history of enhancing emotionality of texts \cite{ScottFahlman,plato}. Research has delved into understanding the emotional states of individuals based on their emoji usage in different settings, including education \cite{Zhang2017_affectivestates}, software development \cite{Chen2021_emotion}, and public forums \cite{Hagen2019_emojiuse}. Findings have also revealed users' innovative ways of utilizing emojis in real life, including re-purposing them from their intended meanings \cite{Wiseman2018_repurposingemoji,kelly2015characterising} or substituting them for text \cite{zhou2017_wechat}. Parallel to academic research, the commercial sector has consistently rolled out new emoji designs \cite{newemoji2023,facebookmessenger,Telegram}. 
Current text messaging applications increasingly support users' customization options \cite{applememoji,googlegboard}, and related research also started probing user authoring of multimodal emoticons \cite{vibemoji}, or enhancing mobile communication with haptic experience \cite{TangibleEmotionResponse,socialTouch}.

Besides emoticons, images stickers or memes are also commonly used in parallel with text messages to communicate emotions, which has been studied or supported by HCI explorations.
For instance, Kim et al. \cite{messageImage} introduced a system that recommends images in line with the message's context to enrich the expressiveness of the conversation. Jiang et al. \cite{jiang2017_GIF} discovered that animated GIFs can evoke intricate emotional responses and enhance nonverbal communication in text messaging. Griggio et al.'s DearBoard \cite{Griggio2021} facilitated the shared customization of image stickers across messaging apps, enabling nonverbal exchanges between close partners.

Moreover, the visual attributes of text, such as font styles \cite{emotype} and motion effects of text \cite{ChatAugmentation}, have also been used to augment text chats. Emotype by Choi and Aizawa \cite{emotype} created emotional fonts that aligned with a chosen emoticon.
Wang et al. \cite{EmotionChatPhysiological} employed physiological data to produce text animations as emotional indicators. Buschek et al. \cite{ChatAugmentation} harnessed physiological data and situational elements to personalize fonts, enriching chat experiences.

Recent studies continue extending the nonverbal affective channels for text messaging by integrating new paralinguistic designs. For instance, Yang et al. \cite{affectiveProfile} used a generative technique to change facial expressions in profile pictures as emotional cues to enhance text-based communication. The EmoBalloon system by Aoki et al. \cite{emoBalloon} makes use of generated explosion-shaped chat bubbles to convey arousal detected from a text message. Although initially designed for text messaging, this study (along with Shi et al.'s work on conversational agent \cite{voiceAgent}) showed the potential of message bubbles, a common component in both text and voice messaging apps, inspiring our exploration of chat bubble animations.


\subsection{Affective Aid in Audio Transcription}
Apart from text messaging, similar paralinguistic components have been explored to aid users' affective comprehension (or communication) with real-time audio transcription. 
For example, Emojilization by Hu et al. \cite{emojilization} is a Speech-to-Text technique that translates emotions from speech into emojis. As related, Zhang et al.'s Voicemoji offers a voice-based emoji entry technique developed for people with visual impairments.
Oomori et al. \cite{DDHEmoji} implemented an emoji-based captioning system to support deaf or hard-of-hearing (DHH) individuals in voice-only meetings.
Animated text was also explored to support TV audiences with hearing impairment \cite{emotionalSubtitles}.
Similarly targeting the DHH community, 
Kim et al. \cite{cpationVisPara} visualized pitch and other nuanced paralinguistic cues using the caption font elements;
Alonzo et al. \cite{CaptionVisNonSpeech} focused on recognizing non-speech sounds and conveying them via text or graphic captions;
and de Lacerda Pataca et al. \cite{captionVis} developed captioning techniques to convey speech prosody and emotions.
Chen et al. \cite{chen2022designing} found that combining text background color and typography could desirably enhance the emotion and content delivery of Speech-to-Text systems. 

Above studies have shown various benefits of designing paralinguistic affective cues to accompany text messages or real-time audio transcripts. However, little research has delved into the emotional teaser features of pending voice messages.

\subsection{Mobile Voice Message Interactions}
Haas et al. \cite{UsageVoiceMessaging} studied the increasing adoption of voice message interaction and found that it granted convenience and efficiency for communication and enabled asynchronous voice exchange as preferred by many users. 
However, it also shifts some effort from the senders to the receivers and imposes situational constraints that can make receivers postpone message retrieval: e.g., busyness, noisy surroundings, or social concerns \cite{UsageVoiceMessaging}. 
Unlike glanceable visual content, voice messages can not be previewed or skimmed other than fully retrieved \cite{UsageVoiceMessaging}: via audio or reading through the transcript.

This substantiates a need for designing pre-retrieval cues that help users quickly gauge the affective tones and set up emotional anticipation before the appropriate moment to fully dive into the message. HCI research has just started exploring such ``emotional teasers'' as pioneered by Chen et al.'s work on employing colored message bubbles \cite{bubbleColoring}. They used colors like orange, red, grey, and blue to match excited, angry, sad, or serene moods and showed their value in signifying or intensifying emotions in voice messages\cite{bubbleColoring}. They also discovered challenges in using colors such as individual perception differences, and potential conflicts with default bubble colors \cite{bubbleColoring}, which inspired us to explore affective animations of bubbles to complement the color approach. 

Other than Chen et al., HCI research scarcely addressed emotional teaser features of voice messages. However, a series of works innovated other aspects of voice message interactions. For instance, Haas et al. explored the augmentation of voice messages with soundscapes, voice changers, and sound stickers \cite{augmentedVoiceMessage}. MeowPlayLive by Ahn et al. \cite{MeowPlayLive} enriched viewer-animal interaction in live streaming via voice messages. Yang et al.'s ProxiTalk eases the input of voice messages via user intention detection \cite{proxyTalk}.

\subsection{Social-Emotional Interactions on Smartwatches}
While the emotional teaser feature remains an unaddressed opportunity for smartwatch-based voice messaging, research encompassed other forms of social-emotional interactions on smartwatches. 
For instance, Graham-Knight et al. \cite{hapticNotificationSmartwatch} employed smartwatch haptic cues as communication means between intimate users. 
Similarly, studies explored augmented social touch using diverse wearable or on-body methods: \eg, \cite{VisualTouch,Poketouch,Nakanishietal2014_handshaking,Wang2012_keepintouch,Zhang2021_SansTouch}.
ThermalWear \cite{ThermalWear} probed wearable thermal feedback as special assistance for affective comprehension of speech.
additionally, many studies innovated (text) input methods on smartwatches, potentially easing the communication experience \cite{TapNShake,GesturalText,magTouch,crownboard,wristband,watchWriter}.

Significantly, animations are recognized as an intuitive and desirable means to communicate emotional elements in smartwatch-based social interactions. Liu et al. presented Animo \cite{AnimoLiu}, a smartwatch system that mobilized abstract affective animations with basic shapes for users to convey their emotional states based on their physiological data. Significant Otter also by Liu et al. \cite{Otter}, features delicate animations of two otters as the visual medium for close partners to exchange bio-signals and convey intimacy.

Animations have long been recognized as an emotive medium \cite{thomas1995illusion,louToLife,lasseter1987principles}, and also utilized as an expressive channel for various HCI systems to depict or evoke emotions, such as generic user interfaces \cite{chevalier2016animations,Harrison2011}, data visualization \cite{de2017taxonomy,emordle}, mobile messaging \cite{vibemoji}, conversational agent \cite{voiceAgent}, or data-driven storytelling \cite{kineticharts}. Tools like AniSAM \cite{Sonderegger2016_AniSAMAniAvatar} or PrEmo \cite{prEmo} employ short animations as self-report measures to effectively capture users' vivid emotional responses, affording depth beyond what static media can offer \cite{normanPremo}. Despite the established role of animations in emotional expression, their potential as emotional teasers embedded in smartwatch voice message bubbles remains unexplored, driving our investigation interest.

\begin{figure*}[tb]
  \centering
  \includegraphics[width=\linewidth]{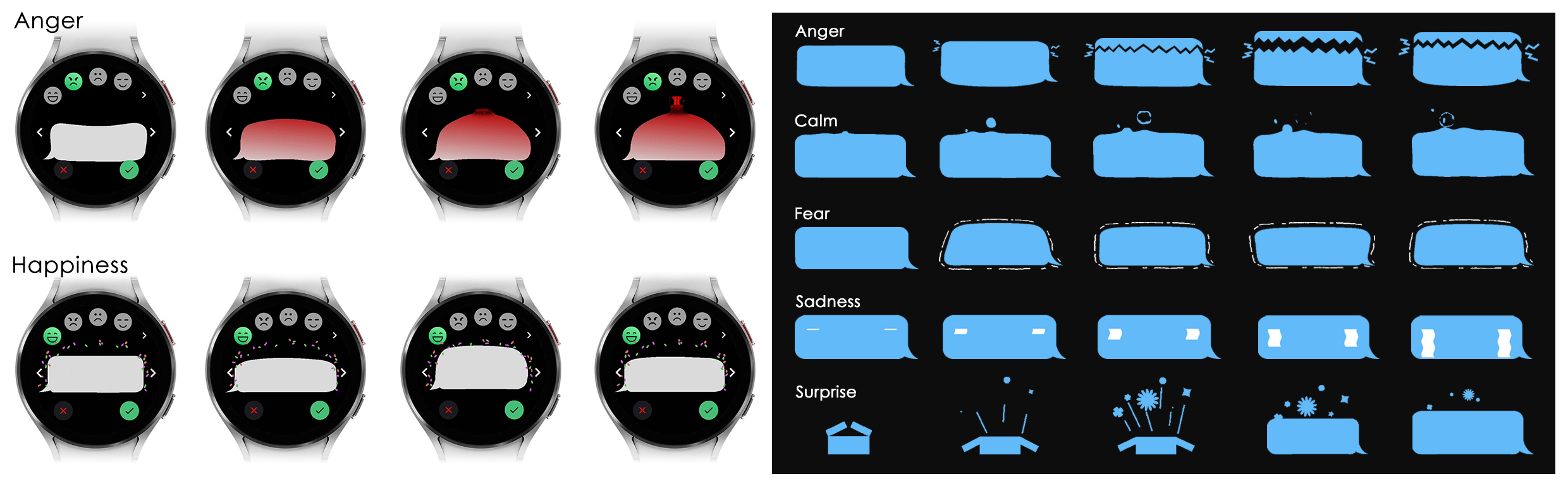}

  \vspace{-4mm}
  \caption{Seven examples from the 30 emotional teaser animations of \name{}.}
  \Description{alt-texts TBA}
  \label{fig:design}
\end{figure*}

\section{\name{} Interface Design}
\subsection{Design Considerations}
The design rationales of the \name{} interface stem from our research objective. HCI research has scarcely explored emotional teaser features for voice messaging on smartwatches. There is little empirical knowledge about how users experience these emotional teasers. Our goal with the \name{} system is to investigate the use of message bubble animations as emotional teasers for voice messages and to understand user experiences with such features, in order to gather opportunities and implications for future research and design. To this end, we formulated a set of design considerations to guide the development of \name{}:

\textbf{D1: Enabling emotion teasing instead of content revealing.} 
Current smartwatch-based voice messages do not support pre-retrieval affective cues for users to gauge the emotional tone before diving into the message.
Yet, users still cherish these messages for the authentic connection they feel when hearing familiar voices \cite{UsageVoiceMessaging}.
To preserve and enhance this genuine bond, our design goal is to facilitate the shared emotional experience and anticipation-building, instead of disclosing the actual content (e.g., akin to the excitement and anticipation before unwrapping a gift rather than spoiling what is inside). As a result, \name{} uses message bubble animations as pre-retrieval emotional cues instead of offering content-based previews like summative transcripts or word clouds.

\textbf{D2: Supporting intuitive and brief engagement.} 
Given that many voice message interactions take place when users are in motion or multitasking \cite{UsageVoiceMessaging}, and considering the limited space on watch screens, the emotional teasers need to be glanceable for quickly and effortlessly discerning emotional tones or building anticipation. To this end, each message bubble animation is designed to continually loop with a short repetition cycle of under four seconds, using straightforward and relatively abstract graphics without intricate patterns or textures. This could ensure intuitive comprehension and avoid prolonged engagement.

\textbf{D3: Balancing convenient selection with diverse options.} 
To enhance the emotional expressivity of voice messages, the system should offer a rich set of animations as teaser options. However, to prevent overwhelming users, a recommendation mechanism is necessary. To balance the convenient selection with diverse options, 
we utilized 30 diverse message bubble animations spanning six emotions \cite{aniBalloon} and developed a fusion model to detect the emotion of an input message based on its semantic and vocal attributes. The interface then highlights two probable emotion categories upfront. While this guides users towards a quick selection, they can still browse all options, tailoring their choice to the specific context. This ensures convenience with user flexibility and autonomy.

\subsection{\name{} Interface and Usage Scenario}
The \name{} interface, illustrated in \autoref{fig:teaser}, enables users to send and receive voice messages with an emotional touch. Users initiate recording by pressing and holding the record button. Upon release, the system processes the audio and prompts the option to append an emotional teaser. In the teaser selection interface, the top section showcases six emojis, each symbolizing an emotion: happiness, sadness, surprise, calmness, fear, and anger. The recommendation algorithm, analyzing both the message's semantic content and vocal tone, determines the initial two emotions displayed. The bottom section lets users preview and select a corresponding bubble animation for their chosen emotion. After selecting, users confirm with a green checkmark button to dispatch the message. Receivers are greeted with the looping animated emotional teaser and can tap it to access the full audio message.

Here we exemplify a typical usage scenario of \name{}.
\re{Alice is at a usual get-together with her friends. Amidst the lively chatter and laughter, her smartwatch buzzes. It's a voice message from her partner, Bob, who's visiting his home country. A glance at the animated message bubble on her smartwatch hints that Bob's message is filled with joy, likely about his experiences back home (\textbf{D2}). Alice, not wanting to disrupt the ongoing conversation with her friends, decides to listen later. The emotional teaser, however, sparks her curiosity and excitement---as she socializes, part of her mind wonders about the joyful story Bob wants to share (\textbf{D1}). After a while, when the group's energy shifts to a more relaxed mode, Alice finds a quiet and laid-back moment to tap on the bubble and listen to the message, with the animation reiterating Bob's emotion. As anticipated, it is about an unexpected and delightful encounter with a childhood friend. Feeling Bob's elation, Alice is eager to record her voice response. After recording, the \name{} interface offers her the option of adding an animated bubble for her message. Using the top bar on the interface, she could indicate the emotion she wants to convey by choosing one of the six icons representing: happiness, sadness, calmness, fear, surprise, and anger. The first two, happiness and surprise, are brought upfront by the \name{} interface according to the content and tone of Alice's message (\textbf{D3}). Alice selects the surprise icon and then chooses from five different bubble animations that best convey her reaction. Satisfied with her choice, she sends the message back to Bob, who can now see her voice message bubble augmented by a teaser animation on his smartwatch.}


\subsection{Message Bubble Animations}

The 30 emotional animations for message bubbles have developed from the AniBalloons project \cite{aniBalloon}. 
Grounded in Ekman's foundational theories on basic emotions \cite{ekman1992argument,ekman1992there}, the design centered on primary emotions: happiness, anger, fear, surprise, and sadness, along with the neutral state of calmness. 
The entire design process spanned a year, adopting the structured affective design method from Kineticharts \cite{kineticharts}. 
The team first curated 230 emotional motion graphics as design inspirations. These were then analyzed to extract animation patterns, emphasizing object motion, dynamic decorative effects, and timing.
These patterns were adapted to generic message bubbles while preserving expressiveness. Expert reviews with professional animation designers led to further refinements, resulting in five distinct animations for each emotion category. 
As detailed in \autoref{fig:design} and the \textit{supplementary materials}, for instance, anger animations convey tension with effects like fire-blowing and body-clenching. Calmness designs evoke serenity using metaphors of water, air, and floating. 
Fear is captured through trembling motions and fluctuating silhouettes, while happiness is shown with celebratory jumps and dancing. Sadness is portrayed through tearful motions and melting effects, and surprise utilizes sudden appearances or splashing effects.

To evaluate the 30 message bubble animations, 40 participants (aged 18-44; 42.5\% self-identified women \re{and 57.5\% self-identified men}) were recruited for an \textit{Affect Recognizability} test. The goal was to determine if participants could discern the emotion each bubble animation aimed to convey, without relying on message content (\textbf{D1}). Using a web-based survey, participants viewed the animations in random order. For each, they identified the emotion they believed the sender intended to convey from six target emotions plus an additional``Other'' option with a user input field. This evaluation technique mirrors that used in \cite{kineticharts}. The results confirmed that 80\% of the designs (24 out of 30) were identified by the majority of participants without a hint from message content (a recognition rate above 50\% is considered the benchmark for effective affective design as per \cite{kineticharts,Ma2012GuidelinesFD}). Twenty designs surpassed a 65\% recognition rate. Binomial tests further revealed that for 93\% of the designs (28 out of 30), the intended emotion was the only option recognized at a rate significantly above mere chance (p<.001), while other options did not deviate significantly from random selection (see detailed results in \textit{supplementary materials}).
This highlights a clear consensus among participants about the conveyed emotion.

\section{EmoWear System Implementation}
Using \name{} both as a novel system to study about and a research tool to study with, our aim is to empirically explore user experiences with emotional teasers for voice messages on smartwatches. in doing so, we implemented the \name{} system into a functional prototype, enabling smartwatch users to send and receive voice messages with animated emotional teasers. In this section, we report our development of the back-end fusion model which detects emotions from an input voice message based on its semantic and vocal features. Subsequently, we present an overview of the front and back-end architecture of the whole system.

\subsection{Emotion Classification Algorithm}

As previously noted, \name{} offers a rich set of bubble animations based on six types of emotions: happiness, anger, fear, surprise, sadness, and calmness. In the back-end of \name{}, we aim to develop an algorithm that can detect the sender's emotions from both the semantic and acoustic features of a voice message and bring two probable emotions upfront to ease convenient selection (\textbf{D3}). In conversational emotion recognition tasks, single-modality models, whether relying solely on text or audio, each can face their own limitations due to absent information from the other modality. Furthermore, existing multimodal models such as DANN \cite{lian20b_interspeech}, SPECTRA \cite{yu-etal-2023-speech}, and M2FNet \cite{10096370}, cannot cover all the basic emotion types targeted by the \name{} system. 

As a result, we have adopted a multi-modal fusion approach that combines two pre-trained single-modality models and fuses the Text-to-Emotion and Speech-to-Emotion results at the decision level (see \autoref{fig:system}). One of the advantages of decision-level fusion is that it can fuse data in different formats. Each modality can use its best classifiers or models to extract features and classifications \cite{XU2020347}. In the process of extracting textual modality input from user's voice messages, we employed the Google Cloud Speech-to-Text API. Below are more details about the pre-trained models we have utilized and their fusion framework.

\subsubsection{Speech-to-Emotion model}

Inspired by the good performance of convolutional neural networks (CNN) in speech recognition, the emotion classification model of speech is based on CNN and dense layers. We adopted the model proposed by Marco et al. \cite{mel9122698}, which uses the Mel-frequency Cepstral Coefficients (MFCC) \cite{mfcc1163420} as the feature to train. MFCC is widely used in speech recognition systems because it could represent the amplitude spectrum of the sound wave in a compact vectorial form. In this work, 40 features were extracted for each audio file. 

The dataset used to train this model is the Ryerson Audio-Visual Database of Emotional Speech and Song (RAVDESS) \cite{ravdess}. 67\% of the RAVDESS dataset was randomly selected for model training, utilizing the rest for model testing. We retrained the model using the six emotion labels targeted by \name{}. The highest emotion prediction precision achieved by the trained model was 0.82, \re{while both the average precision and the average recall were 0.7}. We have thereby saved and used this model to get the emotion prediction vector value as the prediction result for the speech (audio) modality.

\subsubsection{Text-to-Emotion model}

To process the semantic feature of the input voice message, we have employed Google’s work on the GoEmotion \cite{goemotions}. GoEmotion categorizes emotions from textual content into 27 types and also offers a heatmap correlation analysis to reveal a hierarchical structure of these emotions \cite{goemotions}. This analysis shows how these 27 emotions relate to and can be condensed into basic emotional types. Leveraging this hierarchical structure, we mapped the relevant emotional labels back to the six emotion classes targeted by the EmoWear system. We fine-tuned the BERT pre-training model using the converted 6-class GoEmotion dataset based on the mapping relationship. 67\% of the GoEmotion dataset was randomly selected for training and the rest of the data was for testing. The trained model achieved an average precision of 0.664 and an average recall of 0.669. Using this model, the emotion prediction and its probability values based on the text modality (semantic features) can be obtained.

\subsubsection{Multi-modal fusion}

The speech-based probabilities $p_s$ and the text-based probabilities $p_t$ for the same utterance are combined as fused probability $p_f$:
$$p_f = w_1*p_s + w_2*p_t$$

$w_1$ and $w_2$ are the fixed weights assigned to the speech and text modalities. The weights determine the degree of contribution of each data modality to the fused probability. 
Given that each single-modality model excels in its corresponding testing dataset but underperforms in an external dataset, we opted for the third-party Multimodal EmotionLines Dataset (MELD) to explore the fusion weights and evaluate the fusion model. 
We tested three conventional fusion ratios: w1:w2 = 1:2, 1:1, and 2:1. While the performance differences among these ratios were slight (= 55.62\%, 54.6\%, and 54.39\%), the ratio of w1:w2 = 1:2 yielded the highest accuracy. We thereby adopted this ratio for the multimodal fusion.
In our tests on the MELD dataset, this fusion model outperformed both the text-based BERT model and the speech-based CNN model in accuracy, underscoring the practicality of the fusion model approach in an unfamiliar dataset \cite{XU2020347}.




\subsection{Overview of System Architecture}

\begin{figure*}[tb]
  \centering
  \includegraphics[width=\linewidth]{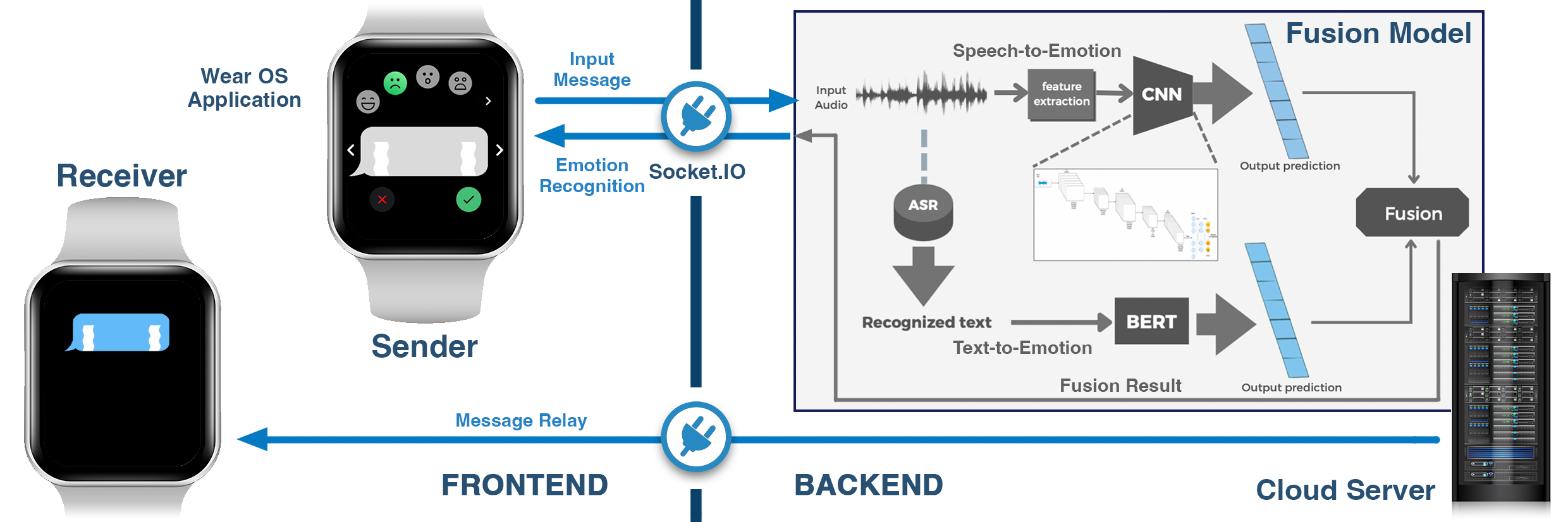}

  \vspace{-4mm}
  \caption{Overview of the \name{} system architecture }
  \Description{alt-texts TBA}
  \label{fig:system}
\end{figure*}

In this section, we present the overall architecture of the functional \name{} system. As shown in \autoref{fig:system}, the system contains a front-end \texttt{Android Wear OS}-based application to support smartwatch users to send and receive voice messages with emotional teasers, and a back-end server to perform data processing and inter-device message relaying. 

The front-end application was built with native Android components based on the Wear OS framework, which communicates with the back-end through the network communication handled by \texttt{Socket.IO}. The application includes all the message bubble animations identified by a unique string to ensure fast local retrieval. The back-end server was developed using \texttt{Node.js}, which is responsible for processing the inputting voice messages and generating emotion predictions for senders, as well as relaying messages between the senders and receivers.

While sending a message, the sender's smartwatch first records an audio message. The message will be immediately sent to the server to be processed for emotion detection. A Speech-to-Text module generates the input for the Text-to-Emotion module of the fusion model, while the audio is handled by the Speech-to-Emotion module. The prediction results generated by the fusion model are then sent back to the sender's device on which two probable emotion categories will be brought upfront to ease user selection. When the user is satisfied with the option and presses to confirm, the audio and the encoding of the selected bubble animation will be relayed to the receiver's device to display the incoming audio message and its bubble animation.


\section{Evaluation Methodology}

Given that HCI research scarcely accumulated empirical knowledge about user experiences of emotional teaser features for voice messages, we utilized the \name{} system as a research tool to probe: How would receivers and senders experience the emotional teasers on smartwatches (\textbf{RQ1})? And what are the relevant design implications for such emotional teasers (\textbf{RQ2})? To empirically address these questions, we employed a mixed-method, within-group comparative evaluation and gathered both quantitative and qualitative results, in order to inform and inspire future design and research.

\subsection{Baseline System for Comparison}
\begin{figure*}[tb]
  \centering
  \includegraphics[width=\linewidth]{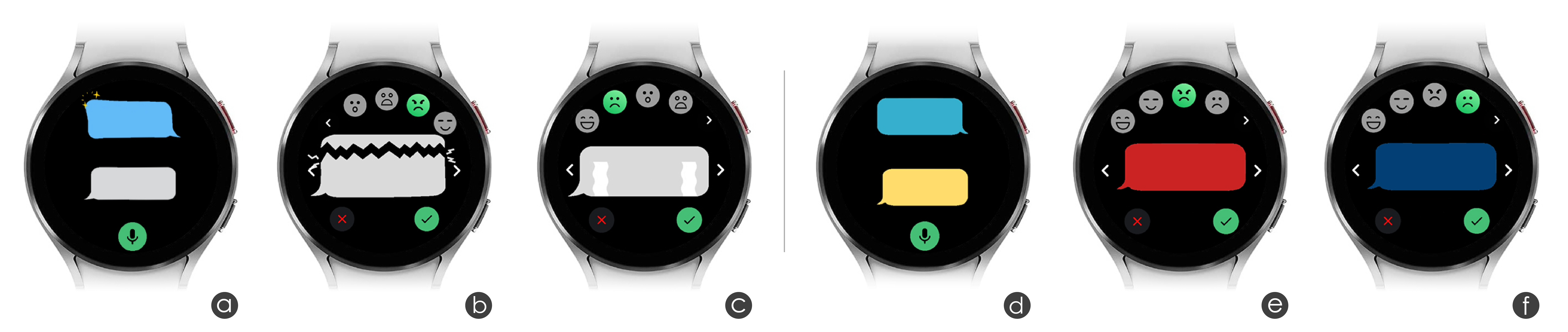}

  \vspace{-4mm}
  \caption{(a,b,c) the \name{} system; (d,e,f) the \baseline{} system implemented for comparison; (a,d) message receiving/recording; (b,e) an example emotional teaser of anger; (c,f) an example emotional teaser of sadness. In \name{}, each emotion has five distinct animations, while in the \baseline{} system, each emotion has five color variations (differing in brightness).}
  \Description{alt-texts TBA}
  \label{fig:baseline}
\end{figure*}

\re{To gather rich insights \cite{GreenbergBuxtonHarmful}, we have compared our EmoWear system with another design variation of emotional teasers: the Baseline system as illustrated in \autoref{fig:baseline}.
We developed the \baseline{} system to mirror the interaction flow and back-end infrastructure of the \name{}.
}
The sole distinction is that the baseline uses message bubble colors as emotion teasers, a method previously explored in mobile voice messaging \cite{bubbleColoring}.
The colors assigned to emotions are grounded in empirical research on color-emotion association \cite{HanadaColourEmotion,ColourPerception,ColourPerception} and previous attempts to visualize emotions through colors in remote communication \cite{ErtayEmoVis}. While consensus exists for some pairings like anger with red and joy with yellow, others like fear, surprise, sadness, and calmness lack clear agreement. 
Following the established practices above and aiming for distinct, harmonious message bubbles without color overlap for the six target emotions, we settled on these emotion-color pairings: anger with red, joy with yellow, fear with purple, surprise with dark cyan, sadness with dark blue, and calmness with light blue.
To mirror the multiple options offered by \name{} under each emotion, in the baseline system, we introduced variations in the colored bubbles, adjusting their brightness to express different levels of the respective emotion. 
\re{It is essential to clarify that our comparison is not aimed for concluding superiority of color or animation approach in general---which is also not useful for design, since color can serve as part of animation. Instead, comparing the two specific design variations affords us to more broadly probe user experiences with emotional teasers, enabling future work to devise new emotional teaser designs inspired by our empirical findings.}

\subsection{Participants and Study Procedure}

\subsubsection{Participants}

We recruited 24 participants (referred to as P1 - 24) through an institute mailing list and snowball sampling ($11$ self-identified males, $13$ self-identified females; aged $21-39, M=25, SD=4$). 
They were invited to participate in the study in person and each session lasted approximately one hour, after which each participant was remunerated with \$15 
The study was approved by [Institute anonymized for review]'s ethics review board.

\subsubsection{Procedure}
Two participants were paired for each study session, where they tested both the \name{} system and its counterpart baseline system in two randomized rounds. Among the 12 pairs of participants, 8 pairs knew each other before the study. They were friends, roommates, or romantic partners. The other four pairs of participants were paired randomly. Upon arrival, they were introduced to the systems and given a smartwatch each (Samsung Galaxy Watch 4) loaded with the two systems (\name{} and \baseline{}). 
They were then separated into different rooms for the testing phase. In the first round, participants interacted with one of the systems, either using animated bubbles or colored bubbles, determined randomly. After using the system, they completed a survey detailing their experiences. The second round allowed participants to test the other system they had not interacted with in the first round. This was followed by a similar survey. Once both rounds were completed, participants were invited for a semi-structured interview to share their insights. Both participants were in the same room during the interview session and they were encouraged to discuss with each other as they responded to the questions.

Each testing round comprised two tasks: a \textit{Scripted Conversation} and an \textit{Unstructured Conversation}. For the \textit{Scripted Conversation}, participants were given conversation scripts, in which they were asked to enact both the line of the conversation and the emotion behind the line. These scripts, adapted ESL Fast \footnote{https://www.eslfast.com/}, contained examples of everyday conversations (see the scripts in the \textit{supplementary documents}) featuring both pronounced emotional expressions and ease of reading. In each testing round, participants needed to complete two sets of scripts, each of which contained four lines that took around 5 minutes to complete. For the second task, participants engaged in an \textit{Unstructured Conversation} for 5 minutes. several common topics were offered as mere suggestions to prompt participants' natural dialogue, such as updating each other on recent events, discussing current news, or planning future activities. Participants were made aware that they were free to discuss any topic of their choosing.


\begin{table*}[tb]
  \caption{Quantitative measures (using seven-point Likert scales).}
  \vspace{-3mm}
  \centering
  \label{tab:measures}
  \small
  \begin{tabular}{llp{0.8\linewidth}}
    \toprule
    
    \textbf{SQ1} & \textbf{R1.1} & Before listening to the audio message, I could tell the emotion my partner wants to express by looking at the bubbles.\\
    & \textbf{R1.2} & In general, after listening to the message, the emotion my partner wants to express closely aligns to my initial guess upon seeing the bubbles.\\
    & \textbf{R1.3} & The bubbles helped me understand my partner’s emotions.\\
    & \textbf{R1.4} & The first time I see the bubbles, I could tell the emotion they represent immediately.\\
    \midrule
    \textbf{SQ2}& \textbf{R2.1}  & I think my emotions were clear to my partner.\\
    & \textbf{R2.2} & The bubbles helped me convey my emotions.\\
    & \textbf{R2.3} & The bubbles could capture the subtleties and nuances of my emotions.\\
    & \textbf{R2.4} & Having multiple options under each emotion tag helps me express my feelings more accurately\\
    \midrule
    \textbf{SQ3} & \textbf{R3.1} & Our conversation was lively.\\
    & \textbf{R3.2} & Our conversation was expressive.\\
    & \textbf{R3.3} & Some nonverbal information supported our communication.\\
    & \textbf{R3.4} &  Inclusion of Other in the Self (IOS) Scale \cite{aron1992inclusion}\\
    \midrule
    \textbf{SQ4} 
    & \textbf{R4.1} & I would like to use it in my life to communicate with friends/family/other people \\
    & \textbf{R4.2} &  I enjoy using the system and I think it’s fun.\\
    & \textbf{R4.3} &  I think the system is easy to use.\\
  \bottomrule
\end{tabular}
\end{table*}

\subsection{Measurements}
After each round, the participants—each experienced the system as both a sender and receiver—filled in a questionnaire. The questionnaire, as shown in \autoref{tab:measures}, comprised fifteen Likert scales tailored to address \textbf{RQ1} and encompassed various experiential qualities of emotional teasers. Namely, these rating scales were clustered to explore the following sub-questions (\textbf{SQ1-6}):

\textbf{SQ1: Can emotional teasers help receivers interpret the emotions of senders?} Four items under \textbf{SQ1} were used to probe the receiver's point of view both pre and post-audio playback: receivers' interpretation of the sender's emotion prior to hearing the audio (\textbf{R1.1}), the perceived congruence between their pre-retrieval interpretation and the post-retrival understanding (\textbf{R1.2}), their general grasp of the sender's emotions (\textbf{R1.3}), and the intuitiveness of the emotional teasers (\textbf{R1.4}, also aligning to \textbf{D2}).

\textbf{SQ2: Can emotional teasers help senders express their emotions?} Four items examined the sender's viewpoint: perceived clarity in conveying emotions to receivers (\textbf{R2.1}), the extent to which the emotional teaser aided their expression (\textbf{R2.2}), the subtlety and nuance in the emotional conveyance (\textbf{R2.3}), and the benefit of having multiple options for each emotion (\textbf{R2.4}, echoing \textbf{D3}).

\textbf{SQ3: Can emotional teasers enhance the communication experience?} \textbf{SQ3} gauges the conversation's liveliness (\textbf{R3.1}) and expressiveness (\textbf{R3.2}), and perceived support from nonverbal cues (\textbf{R3.3}), and perceived closeness (\textbf{R3.4}). The sense of closeness is measured via Inclusion of Other in the Self (IOS) Scale \cite{aron1992inclusion}.

\textbf{SQ4: What are users' general attitudes toward the system?}
Three items probed users' willingness to use in life (\textbf{R4.1}), the system's fun-of-use (\textbf{R4.2}), and its ease-of-use (\textbf{R4.3}).


\subsection{Qualitative Data Gathering}

Upon the finish of the two-round usage, each participant pair participated in a semi-structured interview aimed at obtaining detailed insights to complement and contextualize the quantitative user experience data (\textbf{RQ1}), and further probe design implications for future work (\textbf{RQ2}). The interview was structured in the following parts: First, participants shared their voice messaging experiences with the two systems under comparison. Subsequently, they detailed their feelings and thoughts when interacting with the system and each other as senders and receivers. In the end, participants envisioned real-world scenarios where the emotional teaser features could be integrated. Throughout the interview, participants were also encouraged to share any additional insights or feedback. All interview sessions were audio-recorded, and the content was transcribed verbatim for a thematic analysis.
\re{
The thematic analysis was conducted inductively, allowing themes emerging from the data. Two researchers collaboratively annotated the data and formulated the coding scheme, and then separately coded the data before a iterative review to refine the coding results collectively. 
Although the analysis was guided by participant responses, several emerging clusters were identified to align with our research questions and design considerations.
Regular team discussions ensured the validity and reliability of the findings.
}

\begin{figure*}[tbh]
  \centering
  \includegraphics[width=\linewidth]{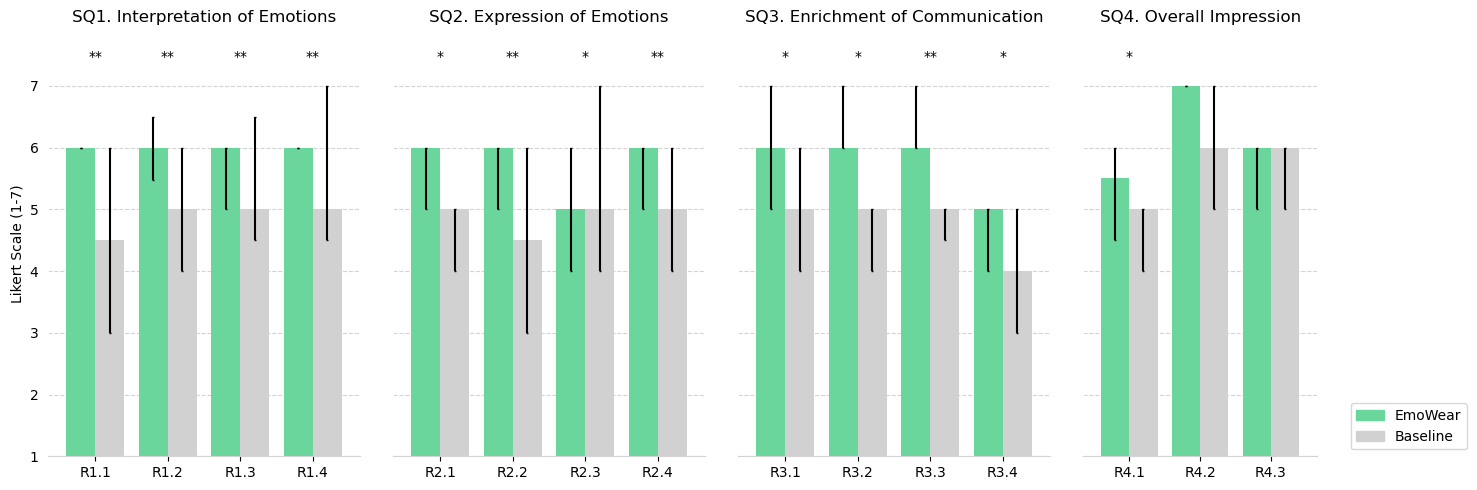}

  \vspace{-4mm}
  \caption{User perceptions of \name{} and \baseline{} on $7$-point Likert scales ($1=$ Strongly Disagree, $7=$ Strongly Agree, $4=$ Neutral; bar lengths represent medians, error bars represent $95\%$ CI by bootstrapping; * $p<=.05$; **$p<=.01$).}
  \Description{alt-texts TBA}
  \label{fig:quant}
\end{figure*}

\section{findings}

\subsection{Quantitative Results}


Below, we present our quantitative findings on different experiential aspects of emotional teasers (\textbf{RQ1}), comparing the \name{} system with the baseline system using $7$-point Likert scale questionnaire data. As depicted in \autoref{fig:quant}, both systems received positive feedback from participants, scoring above the neutral point. However, the \name{} system consistently achieved higher ratings with statistic significance in most (13 out of 15) items, based on Wilcoxon signed-rank tests. Below we report the median value (Mdn), $p$-value, and the effect size ($r$) of each item for comparison.


\subsubsection{Interpretation of Emotions (SQ1)}

Both systems helped receivers interpret the emotions of the sender.
With significant differences, \name{} was considered to better support pre-retrieval prediction of emotions within awaiting messages (\textbf{R1.1}; $\text{Mdn} = 6.0 > 4.5, p = .002, r = -0.619; $ {\footnotesize \text{ i.e. Median$_\text{\name{}} = 6.0 > 4.5 =$ Median$_\text{\baseline{}}$}, \text{$p\text{-value} = .002$}, \text{$r = -0.619$}}), and such prediction was considered to be more aligned with the messaging content after accessing the audio (\textbf{R1.2}; $\text{Mdn} = 6.0 > 5.0, p = .002, r = -0.643$).
Compared to \baseline{}, \name{}'s emotional teasers were considered significantly more useful in aiding the understanding of the other side's emotions (\textbf{R1.3}; $\text{Mdn} = 6.0 > 5.0, p < 0.001, r = -0.804$) and more intuitive to interpret when seen at the first time (\textbf{R1.4}; $\text{Mdn} = 6.0 > 5.0, p = .009, r = -0.533$).
 
\subsubsection{Expression of Emotions (SQ2)}


Both systems facilitated senders' expression of emotions and could capture the subtleties and nuances of the sender's emotions (\textbf{R2.3}; $\text{Mdn} = 5.0 = 5.0, p = .012, r = -0.526$). With significant differences, participants felt that their emotions were displayed more clearly by the emotional teasers of \name{} (\textbf{R2.1}; $\text{Mdn} = 6.0 > 5.0, p = .041, r = -0.417$), and \name{} was more helpful in conveying their emotions than \baseline{} (\textbf{R2.2}; $\text{Mdn} = 6.0 > 4.5, p < 0.001, r = -0.818$).
Users also reported that the multiple options of \name{} helped them more accurately express their feelings than the options of \baseline{} (\textbf{R2.4}; $\text{Mdn} = 6.0 > 5.0, p = .002, r = -0.642$).

\subsubsection{Enrichment of Communication Experience (SQ3)}


Both systems enhanced the affective communication experience for users. With significant differences, the conversations through \name{} were experienced to be more lively (\textbf{R3.1}; $\text{Mdn} = 6.0 > 5.0, p = .022, r = -0.467$), expressive (\textbf{R3.2}; $\text{Mdn} = 6.0 > 5.0, p = .020, r = -0.476$), and better supported by nonverbal information (\textbf{R3.3}; $\text{Mdn} = 6.0 > 5.0, p = .004, r = -0.589$), compared to \baseline{}. They also reported a stronger sense of interpersonal closeness to their partners when using \name{} (\textbf{R3.4}; $\text{Mdn} = 5.0 > 4.0, p = .027, r = -0.451$). 

\subsubsection{Overall Impression of Emotional Teasers (SQ4)}


Users found both systems to be easy to use (\textbf{R4.3}; $\text{Mdn} =  6.0 = 6.0, p = .287, r = -0.217$). 
And they think \name{} and \baseline{} are both fun and enjoyable, with a slight inclination towards \name{}  (\textbf{R4.2}; $\text{Mdn} = 7.0 > 6.0, p = .137, r = -0.303$).
With significance, participants expressed a stronger desire to use \name{} for real life communication than \baseline{} (\textbf{R4.1}; $\text{Mdn} = 5.5 > 5.0, p = .050, r = -0.401$).

\subsection{Qualitative Results}
\subsubsection{\textbf{Contextualizing User Experiences of Emotional Teasers (RQ1)}}
Participants' articulation in interviews helped us complement and contextualize the quantitative results presented above. For instance, supporting the results of sub-question \textbf{SQ1}, the participants reported that the animated emotional teasers aided their pre-retrival interpretation: \qt{So when you see the bubble animation, you kind of get a hint of what the person is thinking. So even before [hearing the audio], you can get an idea of what they're trying to express. So I think it really helps (P11).} And they appreciated the intuitiveness of animations: \qt{animations are intuitive, such as the crying [sadness] and the celebration [happiness] one... (P19)} \qt{if you're happy you can see the thing bounce up and down. (P21)} and \qt{if you're angry, you can see the fires and all that coming out of that (P11)} Such appreciation confirms the value of our design consideration \textbf{D2}.

Senders' point of view addressed by \textbf{SQ2} was also contextualized by the participants verbally: \qt{Animation was nice. I mean, it helped me express my feelings... (P15)} And as (P11) reasoned, compared with static cues, animations are \qt{more vibrant, more vivid and more clear picture [in] describing your emotion...} As a result, participants valued the rich options provided to them: \qt{...really a lot of animations here, which I think will be able to express my emotions more clearly. (P11)} Options can also grant nuances:
\qt{there was more range to show emotion when using the animation versus the colors (P21)} with which \qt{I could express myself in a variety of ways. (P17)}
Meanwhile, Participants also raised the risk of too many options burdening users' minds: \qt{I don't want to spend too much time to make a choice (P10)}. The recommending mechanism was thereby useful: \qt{it's good to have [system recommendations]. Otherwise, you have to choose yourself every time, and that would be tedious (P9)}. Yet still, rich options were strongly favored: \qt{there were a variety of options. There could be more though. (P17)} 
This further underscores the importance of easing users' selection while keeping rich options available to them, as argued by \textbf{D3}.

Participants' comments also confirmed their enhanced communication experience (\textbf{SQ3}). Namely, the emotional teasers served as a nonverbal channel, granting extra bandwidth: \qt{It adds to the density of information. (P7)} Thus it might offload some information from the verbal to the nonverbal:
\qt{the other person will understand me even if it's like a shorter message... (P17)} Their words also helped explain animations' effect on perceived closeness: \qt{so emotional connection is a different level than I feel about connecting with someone in the text... It's more than that. (P11)} And \qt{emotions would really help me understand the person better, without even looking at that person. (P16)} As a result, they consider animated emotional teasers to be suitable for communicating with friends or romantic partners rather than professional scenarios (P7).

As for participants' general attitudes towards the experienced systems (\textbf{SQ4}), they also provided telling accounts for their
willingness to use: \qt{I would definitely use it because it is just easier for me to understand emotions like to be connected to a person at an emotional level. (P16)}
Fun of interaction: \qt{I sort of laugh at the animations themselves [...] So it just like compels me to like, send more messages. (P5)}
And the system' ease-of-use: \qt{it's just easy to like, you know, speak and get it done. (P16)}

\subsubsection{\textbf{Summarizing Desirable Interaction Qualities (RQ2) — Why Emotional Teasers are Considered Helpful and How They Could Support Users}}

To gather implications for future HCI design and research (\textbf{RQ2}), here we summarize the desirable interaction qualities articulated by the participants, surfacing why emotional teasers are considered helpful and how they could support users:

\textbf{Supporting prediction of emotions.} 
Emotional teasers could help users predict and understand the emotional context of an impending message, reducing the uncertainty of the information or the chance of misinterpretation.
For example, as P16 noted, \qt{I would always like to know the other's feeling so that I don't decipher their emotions and their context of saying in a wrong fashion [...] I think animations really helped me understand that part [...]}  
As P1 felt, with emotional teasers, \qt{I could understand what was possibly being portrayed before I listened to the audio.} 
Similarly, \qt{before listening to the audio, we're able to get a gist of the emotion that's conveyed (P22).}

\textbf{Facilitating emotion regulation.} By giving users a glimpse of the emotional tone of an incoming message, emotional teasers can help users prepare for potentially disturbing or surprising information, providing a chance to better self-regulate their emotions. P20 made a telling example on this: \qt{Yeah, and with animation, when he sends the crying animation right away, I can usually tell that he might not be in a good emotional state. So, I mentally prepared myself before opening the message.}

\textbf{Building up curiosity or anticipation} By revealing the emotional tone without disclosing the message content, emotional teasers could create anticipation and curiosity, leading to heightened interest and engagement. As P13 experienced, the emotional teaser \qt{is quite entertaining. And I do enjoy and I quite anticipate the messages coming to me.} Similarly, P5 built up curiosity due to the emotional teasers: \qt{So it feels nice to see the message coming up. And then I want to know what she's saying with that animation.} These comments offered contextual verification for our design consideration \textbf{D1}.

\re{\textbf{Creating a nonverbal emotional atmosphere.}
Emotional teasers might establish an emotional tone or atmosphere for the conversation in parallel with the verbal messages, without requiring extra words: e.g., \qt{I tend to be straightforward and talk about things directly, but (using emotional teasers), I can also convey my emotions (P7).}
P10 mentioned that teaser animations from others could elicit a resonating, or corresponding emotional reactions in her: \qt{when someone is telling me they were really scared after being robbed, and they send those trembling [...] I send the one with an exclamation mark in red, to show that I'm quite shocked [...]}
}

\textbf{Informing decision or prioritization.} 
Emotions could signal the importance, seriousness, or urgency of information, hence guiding the decision-making processes. Emotional teasers could function as such signals: \qt{before listening to the audio, you get a sense of an alert situation (P22).}
This could then help the receiver decide how much priority or attention to give to the incoming message. Namely, as P18 put \qt{if it's a paranoid one, or if it's an angry one, then I know that it's a priority [...] I should read it first [...] if it's a happy mood, then I know that okay, I can read it afterwards.} 

\textbf{Increasing fun and engagement.} As reported earlier, emotional teasers could make digital communication more engaging and fun. As P5 reported, \qt{I feel like it was fun and easy to know her emotions.} Similarly, to P23, \qt{my main reason (for preferring animations) is animations (are) very fun.}
P8 was enthusiastic about creating humorous expressions using emotional teasers: \qt{There are also several types of humor that I can choose to add when it comes to teasing or making fun of people.}

\re{\textbf{Affording expression of closeness or intimacy.}
As shown by the result of the Inclusion of Other in the Self (IOS) Scale, animated emotional teasers increased the perceived closeness between conversational partners. Echoing this, the participants pointed out that families, friends and romantic partners could express intimacy and closeness by using the emotional teasers together. As P7 put, \qt{I think it (emotional teaser) is very suitable for this kind of couple's application, like a couple's watch. This way, I can not only talk to him but also interact with him.}
}

\section{Discussion}

\begin{table*}[tb]
  \caption{Desirable qualities of emotional teasers and contextual opportunities for future HCI design.}
  \vspace{-3mm}
  \centering
  \label{tab:qualities}
  \small
  \begin{tabular}{p{0.3\linewidth} p{0.56\linewidth}}
    \toprule
     \textbf{Desirable Qualities of Emotional Teasers} & \textbf{Contextual Opportunities to Design for} \\
     \midrule
     \textbf{Supporting prediction of emotions} & When users want to ensure the intended sentiment is understood, reduce the emotional uncertainty of communication, or decrease the chance of misinterpretation. \\
     \textbf{Facilitating emotion regulation} & When users need to regulate their emotions or mentally prepare for potentially surprising or disturbing messages. \\
     \textbf{Building up curiosity or anticipation} & When users intend to create suspense or build up curiosity, anticipation, or excitement for an incoming message, akin to the thrill before unwrapping a gift.\\
     \re{\textbf{Creating a nonverbal emotional atmosphere}} & \re{When users desire to set a shared emotional tone or atmosphere for the conversation and trigger mirrored or resonating emotional responses from each other.} \\
     \textbf{Informing decision or prioritization} & When users need emotion to assess the importance, seriousness, or urgency of an awaiting message, and make a decision about how much priority or attention to assign to it. \\
     \textbf{Increasing fun and engagement} & When users aim for more fun, engaging, eye-catching, or humorous experiences in voice message interactions. \\
     \re{\textbf{Affording expression of closeness or intimacy}} & \re{When there is a need to communicate intimacy and closeness, for example, in conversations between families, friends, or romantic partners.} \\
  \bottomrule
\end{tabular}
\end{table*}

Voice messages, unlike glanceable visual content, inherently prevent users from accessing the emotional tone without fully retrieving the audio content \cite{UsageVoiceMessaging}. This limits the experience of shared emotion in the pre-retrieval phase. 
We thereby set out to explore the concept of "Emotional Teasers"—cues that enable users to take a glimpse of the emotional tone in an awaiting message before revealing its content.
\re{
Although HCI research has thoroughly investigated the role of paralinguistic cues in augmenting the emotional depth of text messages and audio transcripts, there is a dearth of knowledge about emotional teasers.
Our study sheds light on the value of providing emotional hint for voice messages before they are accessed. Resonating with Cho et al.'s \cite{ChoSenderControl} research on sender-controlled notifications, our \name{} system highlights the potential of emotional teasers as sender-controlled cues for emotionally enriching communication.
}
Our overall research objective is twofold: first, to empirically understand the user experience of animated emotional teasers in smartwatch-based voice messaging (\textbf{RQ1}); and second, to contextually surface relevant opportunities and implications for future HCI design exploring the feature of emotional teasers (\textbf{RQ2}).

To this end, we have designed and evaluated \name{} as both a novel system to study about and a research tool to study with. \name{} is a smartwatch voice messaging system that allows users to apply 30 animation teasers on message bubbles to reflect senders' emotions. EmoWear eases the sender's selection process by detecting and prioritizing emotions within an input message through semantic and acoustic processing. We compared EmoWear's perceived usefulness with a mirroring system that employs color-coded bubbles to reflect emotions, involving 24 participants (12 pairs) in a within-group study. 
Addressing \textbf{RQ1}, the quantitative results indicate that both \name{} and the \baseline{} systems were positively received, suggesting that participants found emotional teasers generally valuable. Moreover, \name{} outperformed \baseline{} in most metrics (13 out of 15) in Wilcoxon signed-rank tests (see \autoref{fig:quant}).

Furthermore, from the receiver's perspective, \name{} was perceived to enhance emotional interpretation at the pre-retrieval stage, with interpretations aligning more closely with the message content post-retrieval. Additionally, \name{}'s teasers were deemed more intuitive and effective in fostering emotional understanding compared to the baseline. For senders, \name{}'s animated teasers were considered to enable clearer emotional conveyance. The options in \name{} were also appreciated for their potential to facilitate more accurate emotional expression than the options of the baseline. Interestingly, both \name{} and the baseline were similarly and positively rated in capturing the nuances and subtleties of the sender's emotions.
In terms of overall communication experience, \name{} was perceived as more vibrant and expressive, better enhancing non-verbal aspects of communication and fostering a sense of interpersonal closeness. Lastly, while both systems were appreciated for their ease-of-use and fun-to-use, participants expressed a significantly stronger willingness to incorporate \name{} into their daily communication routines.

The qualitative data offered concrete examples and explanations to contextualize above metrics.
Moreover, in response to \textbf{RQ2}, we have summarized a set of desirable interaction qualities of emotional teasers, to further shed light on why emotional teasers could be meaningful and in which way they could support voice message interactions in daily contexts:
\textit{supporting prediction of emotions} to reduce emotional uncertainty or misinterpretation,
\textit{facilitating emotion regulation} by enabling users' self-regulation and mental preparation before accessing surprising or disturbing messages, 
\textit{building up curiosity or anticipation} to create an experience of suspense or excitement without spoiling the message content, 
\textit{enhancing emotional contagion or empathy} to help users set an emotional ambience or evoke empathetic responses, 
\textit{informing decision or prioritization} by aiding quick decisions on how much priority or attention an awaiting message deserves, 
\textit{increasing fun and engagement} to grant enjoyable or humorous experiences in messaging, and 
\textit{conveying intimacy or closeness} in conversations between families, friends, and lovers. 
To further inspire broader explorations on emotional teaser features, we formulate a set of contextual opportunities to inspire future HCI design, each corresponding to a desirable quality discovered in our study, as listed in \autoref{tab:qualities}. Continuing addressing \textbf{RQ2}, below we generalize and discuss a set of design implications to inform future HCI research:

\re{
\textbf{Implication 1: envisaged applicable scenarios of emotional teasers for further exploration.} 
Based on the rich articulations from the participants, several applicable scenarios of emotional teasers have been surfaced where emotional teasers are viewed particularly beneficial. These scenarios could concretely inform future HCI research to broadly expand the application domains of emotional teasers. Namely, these promising scenarios could be summarized into three categories:
(1) \textbf{Occupied-Hands Scenarios:} Participants notably envisioned the value of EmoWear teasers in situations where their hands were busy, such as cooking or driving. In these instances, a brief look at an emotional teaser can give a hint of the message's nature when listening is not immediately possible. (2) \textbf{On-the-Move Scenarios:} Another significant scenario involves users constantly moving, whether indoors or outdoors. Here, the noise and busyness often lead to the postponement of voice message retrieval. An easily visible emotional cue in these moments can be practical, offering a quick glace at the message's tone. (3) \textbf{Social Scenarios:} The third scenario highlighted was during social events, ranging from formal meetings to informal gatherings. In such settings, immediate access to audio messages might be socially unsuitable or could expose privacy. Emotional teasers are thus seen as an handy feature, allowing users to grasp the emotional essence of a message without disrupting their social engagement.
These diverse scenarios provide crucial directions for future HCI research and practice, especially given the nascent stage of emotional teasers in HCI and the existing gap in ecologically understanding their usage and utility in various real-life settings.
}

\textbf{Implication 2: further expanding emotion options and exploring multi-emotion representations in emotional teaser design.} 
Users valued the diverse animated emotional teasers offered by \name{}. Yet, they expressed a desire for more animation options to convey a broader spectrum of emotions, especially those that are complex or nuanced, stemming from basic emotions. For example, participants expressed interest in teasers that could convey emotions like love, sarcasm, worry, doubt, hopefulness, or contemplation. One participant suggested a teaser for "comforting" when a conversation partner is facing challenges or anxiety.
Additionally, participants saw potential in representing multiple emotions within a single voice message. They highlighted scenarios where conveying a shift or transition in emotions within a message might be valuable. Drawing inspiration from prior work on paralinguistic emotional cues for real-time speech transcriptions, such as Emojilization \cite{emojilization} and the study by Oomori et al. \cite{DDHEmoji}, which translate speech emotions into emojis, we suggest future research could explore creating a sequence of animations. These sequences could represent the progression or transition of emotions within a message. However, it's essential to differentiate between designing emotional teasers and paralinguistic emotional cues. Emotional teasers provide a snapshot into the voice message's emotions before accessing the audio, while paralinguistic cues enhance real-time text captions or transcriptions. Thus, if emotional teasers were to depict a sequence or transition of emotions, they should remain concise and easily digestible, allowing users to quickly grasp the message's tone without significant interruptions.

\textbf{Implication 3: leveraging user authoring or (co-) customization for closeness and intimacy.} 
While many users found the animations valuable for intimate conversations with close friends, family, and significant others, they felt it less appropriate for professional or workplace communications. 
The Inclusion of Other in the Self (IOS) Scale \cite{aron1992inclusion} supports the idea that animated emotional teasers can foster closeness and intimacy in voice messaging. This sentiment aligns with Liu et al.'s findings \cite{AnimoLiu,Otter}, suggesting that affective animations on smartwatches can effectively convey closeness or intimacy.
Building on this, we connect our design implication with the work of DearBoard \cite{Griggio2021} by Griggio et al. Their research highlighted the intimacy achieved when users co-customize elements like image stickers or emoticon shortcuts on their shared messaging platforms, enhancing nonverbal exchanges and feelings of connectedness. \re{Moreover, Griggio et al. \cite{GriggioCustomization} have unveiled that customization also contributes to users' expression of identity, besides intimacy to others.}
Translating this to emotional teasers, we envision a future where users with intimate relationships co-create their unique set of emotional teasers, symbolizing deeply personal shared experiences. For instance, a pizza emoji, as identified in Wiseman et al.'s research \cite{Wiseman2018_repurposingemoji}, could represent love between two individuals.
\re{To facilitate this, we suggest a low-barrier user authoring approach (as used in \cite{vibemoji}), allowing users to recombine different emotional teaser components to craft new, meaningful animations. Most animations in \name{} consist of "main body movement" and "dynamic decorative elements." By enabling users to mix and match these elements, and allowing them to configure the pace or range of the motion, we can foster richer, more personalized expressions, enhancing real-life communication.}

\textbf{Implication 4: combining multiple information channels or sensory modalities, and incorporating biosignals.} 
In our study, to gain a concrete understanding of user experiences with emotional teasers, we compared our system, \name{}, to our crafted \baseline{} system that employed color-coded message bubbles as emotional indicators. This color-coding approach, previously explored by Chen et al. \cite{bubbleColoring}, which we consider to be a pioneering exploration into emotional teasers for voice messaging. 
\re{By comparing these two design variations, we were able to broadly probe the user experiences of emotional teasers. 
In general, \name{} received higher ratings than \baseline{}, with its teaser animations seen as more expressive and useful for pre-retrieval emotional indication. 
Nonetheless, participants also noted the advantage of colored bubbles: some colors were universally understood and highly intuitive in representing emotions, e.g., red for anger and yellow for joy. 
They also pointed out the limitations of both animation and color teasers:
animations may contain cultural references, such as symbols from Japanese manga, that may not be well understood across cultural groups. 
Whereas, the color schemes might not be accessible to color-blind users, and the interpretation of some colors can be subjective, varying in emotional representation among individuals.
This suggests that future designs could thoughtfully complement the two approaches for more comprehensive and enriching solutions.}
In such a design, both the base color and animation of the message bubbles could work in symphony, to convey emotions.

Furthermore, we see potential in integrating additional modalities beyond visuals, to elevate the expressiveness and overall experience of emotional teasers. For instance, VibEmoji \cite{vibemoji} combined emoticons, animations, and vibrotactile patterns, allowing users to communicate through multimodal emoticons. Similarly, Haas et al. enhanced voice messages with soundscapes, voice modifiers, and sound stickers \cite{augmentedVoiceMessage}. Drawing from these innovations, we propose that future research could merge multiple information channels or modalities, such as blending visual cues with affective sound effects and vibrotactile patterns. This would provide a richer, more immersive emotional teaser experience.
Additionally, inspired by Liu et al.'s Animo and Significant Otters \cite{AnimoLiu,Otter}, which utilized affective animations to convey biosignals between closely connected users on smartwatches, we believe there's potential in integrating users' biosignals. Such bio-data-driven emotional teasers could offer a more genuine, human-touch experience in asynchronous messaging.

\textbf{\re{Future potential:} improving accessibility and inclusivity of voice messages using emotional teasers.}
Participants raised the potential of emotional teasers to benefit specific user groups, including neurodiverse individuals and the Deaf or Hard-of-Hearing (DHH) community. \re{It's important to note that there is currently no evidence to suggest that emotional teasers are effective in aiding DHH or neurodiverse communities.} \re{Nonetheless}, this feedback underscores the \re{future potential} of viewing emotional teasers through the lens of accessibility and inclusivity.
For instance, neurodiverse individuals, such as those with autism, may face challenges in interpreting emotions from voice tones. Emotional teasers can serve as a visual aid, offering a snapshot of the message's emotional context. This can streamline their communication experience, aiding in both comprehension and response formulation.
Similarly, the DHH community often grapples with the voice or audio-based communications. While transcription services can convert voice to text, the emotional undertones often get lost. A considerable body of prior research has explored assistive paralinguistic cues for real-time speech transcription \cite{DDHEmoji,emotionalSubtitles,cpationVisPara,CaptionVisNonSpeech,captionVis,chen2022designing}:
For example, leveraging animated texts for TV audiences \cite{emotionalSubtitles}, or using visual components of captions to visualize speech prosody and emotions \cite{captionVis}, pitch and other nuanced paralinguistic elements \cite{cpationVisPara}, or non-speech sounds \cite{CaptionVisNonSpeech}.
Drawing inspiration from these, we deem that emotional teasers could be a valuable asset in asynchronous voice communication for the DHH community. 
By providing a visual cue about the emotion behind the message, DHH users can get a more holistic understanding of the message, even if they might miss out on the auditory nuances.
Furthermore, for users with color blindness, the design of teasers can incorporate distinct patterns or animations that don't rely solely on color differentiation.

\re{\textbf{Limitations.} 
As an early exploration of emotional teasers, our work faces certain limitations. The foremost is the constraint of the in-lab evaluation---while it richly revealed user experience of emotional teasers, it was limited in capturing the nuances and convexity of real-world settings. Future studies should incorporate in-the-wild evaluations and longer-term studies to understand sustained usage and to mitigate the novelty effect. Additionally, involving a larger sample size would strengthen the validity of the results.
Concerning our system design, the use of a cloud-based recommendation engine may raise privacy issues, and accidental network latency could affect user experience. Future developments might explore less computationally demanding, locally served machine learning techniques, like MFCC, to overcome these challenges.
The potential for bias in the training data for emotional voice recognition, due to imbalanced labels, is another critical aspect. Furthermore, it is important to note that while emotional teasers aid in conveying and interpreting emotional content, their effectiveness and meaningfulness is highly dependent on the context. The interpretation of these teasers could be influenced by various factors including the identities of the sender and receiver, and the specific situations, time, and place. 
For instance, while participants considered emotional teasers to be suitable for casual conversations with family, friends, or romantic partners to boost the feeling of shared emotions, it's unclear how they would fare in communications involving professional, serious, or negative topics. The impact and usefulness of emotional teasers in such scenarios remain to be explored.
This necessitates a comprehensive approach to understand how emotional teasers are interpreted and used in everyday scenarios, similar to the ecological approaches seen in studies by \cite{ChoSenderControl,GriggioCustomization,Griggio2021}.
}

\section{Conclusion}

While voice messages offer a personal touch, they inherently restrict users from discerning the emotional undertones without fully accessing the audio content. Our exploration of "Emotional Teasers" through the \name{} system has sought to address this opportunity, offering smartwatch users a glimpse into the emotional tone of an awaiting voice message.
Our comparative study, involving 24 participants, demonstrated a positive reception towards the concept of emotional teasers. Notably, \name{} consistently outperformed a baseline system in most metrics, highlighting its perceived value in supporting voice message interaction.
\name{} was perceived to enhance emotional interpretation of receivers at the pre-retrieval stage, with interpretations aligning more closely with the message content post-retrieval. \name{}'s animated teasers were considered to enable clearer emotional conveyance for senders. We also generalized the desirable interaction qualities of emotional teasers, surfacing why emotional teasers are considered meaningful and what contextual opportunities future HCI designs could target.
Looking ahead, we propose several avenues for expanding the scope and impact of emotional teasers in HCI. These include diversifying emotion options and facilitating multi-emotion representations, fostering user customization to enhance closeness and intimacy, integrating multiple information channels or sensory modalities, and incorporating biosignals to create a more authentic and inclusive communication experience. 
Moreover, we emphasize the potential of emotional teasers to augment accessibility for special user groups, thereby fostering a more inclusive and empathetic digital communication landscape.
In conclusion, our exploration of \name{} provides a concrete investigation into the novel and under-explored concept of emotional teasers for voice messaging. It not only offers a novel system but also empirical insights that delineate a promising trajectory for future HCI research to utilize emotional teasers to support asynchronous voice communications.

\begin{acks}
We thank Dr. Wei Li and Dr. Da-Yuan Huang, for their support to our work. We appreciate the contributions made by Ms. Chaoyu Zhang and other partners in the prior work on ``AniBalloons''. Our thanks also go to the reviewers for their invaluable comments that helped us significantly improve this paper. This work is supported in part by the Natural Sciences and Engineering Research Council of Canada (NSERC) Discovery Grant, University of Waterloo WHJIL, and Shenzhen Stability Support (No. 20220815171308001).
\end{acks}

\bibliographystyle{ACM-Reference-Format}
\bibliography{references.bib}

\end{document}